\documentclass[preprint,12pt,authoryear]{elsarticle}

\pdfoutput=1

\usepackage{amssymb}
\usepackage{amsmath}
\usepackage{amssymb}
\usepackage{amsfonts}
\usepackage{amsthm}
\usepackage{subfigure}
\usepackage{xcolor}
\usepackage{units}

\usepackage{amsmath}
\usepackage{amsbsy}
\usepackage{amssymb}
\usepackage{amsthm}
\usepackage{amscd}
\usepackage{amsfonts}
\usepackage{supertabular}
\usepackage{graphics}
\usepackage{graphicx}
\usepackage{verbatim}
\usepackage{subfigure}
\usepackage{epsfig}
\usepackage{xspace}
\usepackage{euscript}
\usepackage{alltt}
\usepackage{boxedminipage}
\usepackage{float}
\usepackage{times}

\newcommand{\mbs}[1]{\boldsymbol{#1}}

\graphicspath{
    {./}
}

\journal{}

\begin{document}

\begin{frontmatter}

\title{Oncotripsy: Targeting cancer cells selectively via resonant harmonic excitation}

\author[label1]{S.~Heyden\corref{cor1}}
\cortext[cor1]{Corresponding author}
\ead{heyden@caltech.edu}
\author[label1]{M.~Ortiz}
\ead{ortiz@caltech.edu}
\address[label1]{Division of Engineering and Applied Science, California Institute of Technology, Pasadena, CA 91125, USA}

\begin{abstract}

We investigate a method of selectively targeting cancer cells by means of ultrasound harmonic excitation at their resonance frequency, which we refer to as {\sl oncotripsy}. The geometric model of the cells takes into account the cytoplasm, nucleus and nucleolus, as well as the plasma membrane and nuclear envelope. Material properties are varied within a pathophysiologically-relevant range. A first modal analysis reveals the existence of a spectral gap between the natural frequencies and, most importantly, resonant growth rates of healthy and cancerous cells. The results of the modal analysis are verified by simulating the fully-nonlinear transient response of healthy and cancerous cells at resonance. The fully nonlinear analysis confirms that cancerous cells can be selectively taken to lysis by the application of carefully tuned ultrasound harmonic excitation while simultaneously leaving healthy cells intact.
\end{abstract}


\end{frontmatter}

\section{Introduction} \label{Sec:Introduction}

In this study, we present numerical calculations that suggest that, by exploiting key differences in mechanical properties between cancerous and normal cells, {\sl oncolysis}, or 'bursting' of cancerous cells, can be induced selectively by means of carefully tuned ultrasound harmonic excitation while simultaneously leaving normal cells intact. We refer to this procedure as {\sl oncotripsy}. Specifically, by studying the vibrational response of cancerous and healthy cells, we find that, by carefully choosing the frequency of the harmonic excitation, lysis of the nucleolus membrane of cancerous cells can be induced selectively and at no risk to the healthy cells.

Numerous studies suggest that aberrations in both cellular morphology and material properties of different cell constituents are indications of various forms of cancerous tissues. For instance, a criterion for malignancy is the size difference between normal nuclei, with an average diameter of $7$ to $9$ microns, and malignant nuclei, which can reach a diameter of over $50$ microns \citep{Berman:2011}. Early studies \citep{Guttman:1935} have shown that the nuclear-nucleolar volume ratios in normal tissues and benign as well as malignant tumors do not differ quantitatively. Nucleoli volumes of normal tissues, however, are found to be significantly smaller than the volume of nucleoli in cancerous tissues \citep{Guttman:1935}. Similarly, the mechanical stiffness of various cell components has been found to vary significantly in healthy and diseased tissues. In \cite{Cross:2007}, the stiffness of live metastatic cancer cells was investigated using atomic force microscopy, showing that cancer cells are more than $80\%$ softer than healthy cells. Other cancer types, including lung, breast and pancreas cancer, display similar stiffness characteristics. Furthermore, using a magnetic tweezer, \cite{Swaminathan:2011} found that cancer cells with the lowest invasion and migratory potential are five times stiffer than cancer cells with the highest potential. Likewise, increasing stiffness of the extracellular matrix (ECM) was reported to promote hepatocellular carcinoma (HCC) cell proliferation, thus being a strong predictor for HCC development \citep{Schrader:2010}. Moreover, enhanced cell contractility due to increased matrix stiffness results in an enhanced transformation of mammary epithelial cells as shown in \cite{Paszek:2005}. Conversely, a decrease in tissue stiffness has been found to impede malignant growth in a murine model of breast cancer \citep{Levental:2009}.

Various experimental techniques have been utilized in order to quantitatively assess the material properties of individual cell constituents in both healthy and diseased tissues. The inhomogeneity in stiffness of the living cell nucleus in normal human osteoblasts has been investigated by \cite{Konno:2013} using a non-invasive sensing system. As shown in that study, the stiffness of the nucleolus is relatively higher compared to that of other nuclear domains \citep{Konno:2013}. Similarly, a difference in mass density between nucleolus and nucleoplasm in the xenopus oocyte nucleus was determined by \cite{Handwerger:2005} by recourse to refractive indices. The elastic modulus of both isolated chromosomes and entire nuclei in epithelial cells are given by \cite{Houchmandzadeh:1997} and \cite{Caille:2002}, respectively. Specifically, \cite{Houchmandzadeh:1997} showed that mitotic chromosomes behave linear elastically up to 200\% extension. Experiments of \cite{Dahl:2004} additionally measured the network elastic modulus of the nuclear envelope, independently of the nucleoplasm, by means of micropipette aspiration, suggesting that the nuclear envelope is much stiffer and stronger than the plasma membranes of cells. In addition, wrinkling phenomena near the entrance of the micropipette were indicative of the solid-like behavior of the envelope.

\cite{Kim:2011} estimated the elastic moduli of both cytoplasm and nucleus of hepatocellular carcinoma cells based on force-displacement curves obtained from atomic force microscopy. In addition, \cite{Zhang:2002} used micropipette aspiration techniques in order to further elucidate the viscoelastic behavior of human hepatocytes and hepatocellular carcinoma cells. Based on their study, \cite{Zhang:2002} concluded that a change in the viscoelastic properties of cancer cells could affect metastasis and tumor cell invasion. The increased compliance of cancerous and pre-cancerous cells was also investigated by \cite{Fuhrmann:2011}, who used atomic force microscopy to determine the mechanical stiffness of normal, metaplastic and dysplastic cells, showing a decrease in Young's modulus from normal to cancerous cells.

The scope of the present work, and the structure of the present paper, are as follows. We begin by defining the geometric model and summarizing the material model and material parameters used in finite-element analyses. Subsequently, the accuracy of the finite-element model is assessed by means of a comparison between numerical and analytical solutions for the eigenmodes of a spherical free-standing cell. We then present eigenfrequencies and eigenmodes of a free-standing ellipsoidal cell, followed by a Bloch wave analysis to model tissue consisting of a periodic arrangement of cells embedded in an extracellular matrix. Finally, resonant growth rates are calculated that reveal that cancerous cells can selectively be targeted by ultrasound harmonic excitation. The transient response at resonance of healthy and cancerous cells is presented in the fully nonlinear range by way of verification and extension of the findings of the harmonic modal analysis. We close with a discussion of results.

\section{Finite element analysis} \label{Sec:FE}

{In this section, we investigate the dynamical response of healthy and cancerous cells under harmonic excitation. We begin by briefly outlining the underlying geometric and material parameters used in our analysis, followed by a verification of the finite element model used for modal analysis. We then calculate the eigenfrequencies and eigenmodes of both free-standing and periodic distributions of cells. In this latter case, we determine the full dispersion relation by means of a standard Bloch wave analysis. Finally, we present resonant growth rates and simulate the transient response of both cancerous and healthy cells excited at resonance in a fully-nonlinear setting by means of implicit dynamics calculations.}

\subsection{Geometry and material parameters}
\label{Sec:Geometry_Material}

\begin{figure}[!h]
\begin{center}
    \includegraphics[width=0.8\textwidth]{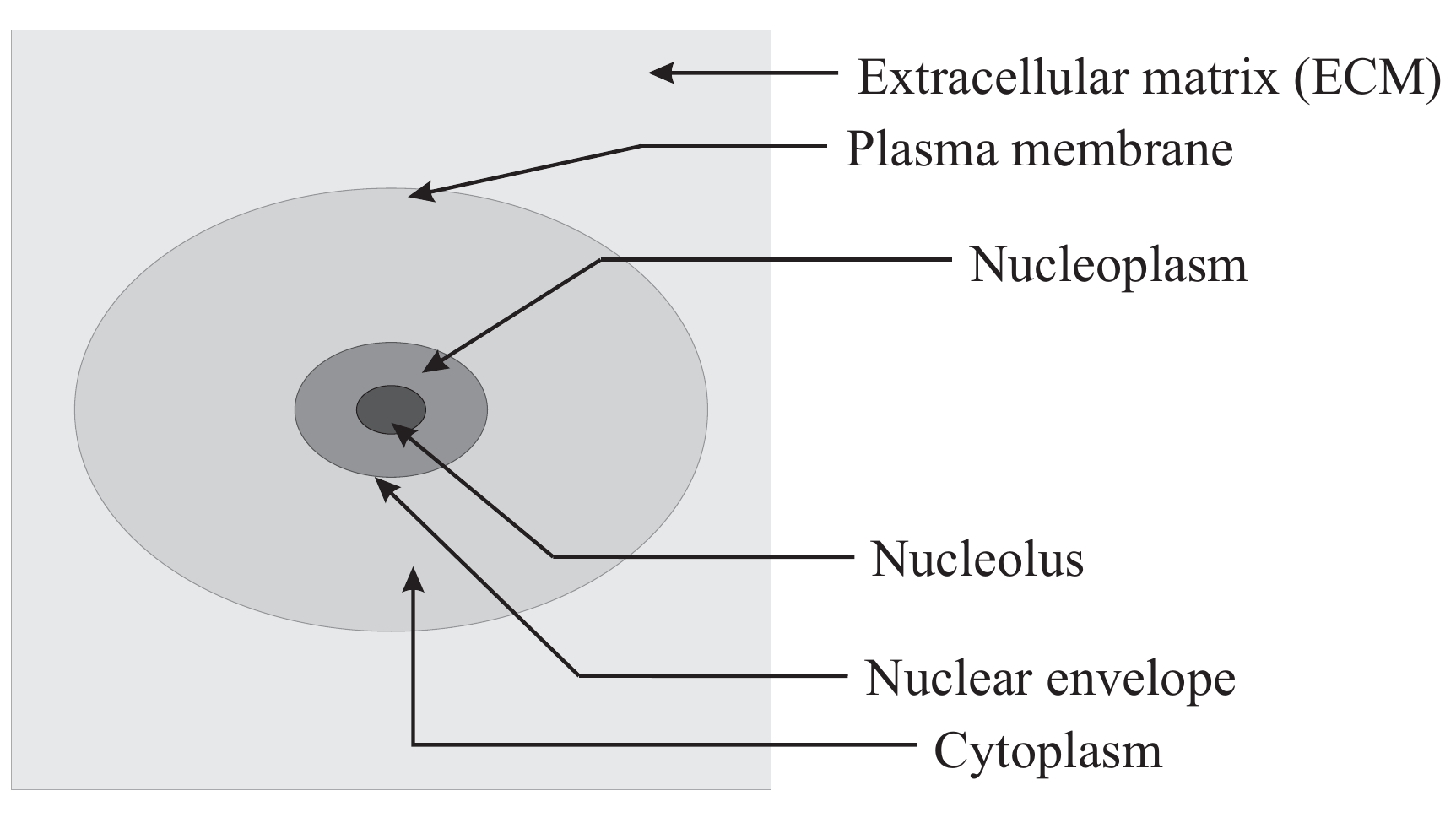}
\caption{Cell geometry with different cell constituents used in finite-element simulations.}
\label{Fig:Geometry}
\end{center}
\end{figure}

The nucleus, the largest cellular organelle, occupies about $10\%$ of the total cell volume in mammalian cells \citep{Lodish:2004, Alberts:2002}. It contains the nucleolus, which is embedded in the nucleoplasm, a viscous solid similar in composition to the cytosol surrounding the nucleus \citep{Clegg:1984}. In this study, the cytosol is modeled in combination with other organelles contained within the plasma membrane, such as mitochondria and plastids, which together form the cytoplasm. For simplicity, we idealize the plasma mebrane, nuclear envelope, cytoplasm, nucleoplasm, and nucleolus as being of spheroidal shape. We model the plasma membrane, a lipid bilayer composed of two regular layers of lipid molecules, in combination with the actin cytoskeleton providing mechanical strength as a membrane with a thickness of $10$\,nm \citep{Hine:1999}. Similarly, we model the nuclear envelope, a double lipid bilayer membrane, in combination with the nuclear lamin meshwork lending it structural support as a $20$\,nm thick membrane. We define the cytoplasm, nucleoplasm, and nucleolus as spheres with radii of $5.8\,\mu$m, $2.7\,\mu$m, and $0.9\,\mu$m and subsequently scale them by a factor of $1.2$ in two dimensions in order to obtain the desired spheroidal shape. We assume an average nuclear diameter of about $5\,\mu$m, as reported in \cite{Cooper:2000}. Diameters for both cytoplasm and nucleolus follow from \cite{Lodish:2004} and \cite{Guttman:1935}, who report nucleus-to-cell and nucleus-to-nucleolus volume ratios of $0.1$ and $30.0$, respectively. The geometry with all cell constituents as used in subsequent finite element analyses is illustrated in Figure~\ref{Fig:Geometry}. In order to further elucidate the effect of an increasing nucleus-to-cell volume ratio, as observed experimentally \citep{Berman:2011}, we consider a range of geometries with increasing nuclear and nucleolar volumes. For all of these geometries, we hold fixed the volume of the cytoplasm. Furthermore, we assume a constant nuclear-to-nucleolar volume ratio for both healthy and cancerous cells, as observed by \cite{Guttman:1935}.

Cell-to-cell differences and experimental uncertainties notwithstanding, the preponderance of the observational evidence suggests that the cytoplasm, nucleus and nucleolus are ordered in the sense of increasing stiffness. Neglecting viscous effects, we model the elasticity of the different cell constituents by means of the Mooney-Rivlin-type strain energy density of the form
\begin{equation}
    W(\mbs{F})
    =	
    \frac{1}{2}\left[\mu_1\left(\frac{I_1}{J^{2/3}}-3\right)
    +
    \mu_2\left(\frac{I_2}{J^{4/3}}\right)
    +
    \kappa\left(J-1\right)^2\right] ,
\end{equation}
where $\mbs{F}$ denotes the deformation gradient, $J = \det(\mbs{F})$ is the Jacobian of the deformation, and $\mu_1$, $\mu_2$ and $\kappa$ are material parameters. For both cytoplasm and nucleus in cancerous cells, material parameters corresponding to the data reported by \cite{Kim:2011} are chosen and summarized in Table~\ref{Tab:Parameters}. We additionally infer the elastic moduli of the nucleolus from \cite{Konno:2013} based on a comparison of the relative stiffnesses of the nucleoli and other nuclear domains. For membrane elements of the plasma membrane and nuclear envelope, we choose material parameters corresponding to the cytoplasm and nucleoplasm, respectively. Furthermore, we infer matrix parameters from the shear moduli reported by \cite{Schrader:2010} for normal and fibrotic livers. For all parameters, we resort to small-strain elastic moduli conversions, with a Poisson's ratio of $0.49$ to simulate a nearly incompressible material, in order to match experimental values with constitutive parameters. We vary the stiffness of both cellular components and extra-cellular matrix (ECM) within a pathophysiologically-relevant range in order to investigate the effect of cell softening and ECM stiffening on eigenfrequencies. Finally, we assume both cytoplasm and nucleoplasm to have a mass density of $1$\,g/cm$^3$, a value reported by \cite{Moran:2010} as an average cell density, and we set the density of the nucleolus to $2$\,g/cm$^3$ \citep{Birnie:1976}.

\begin{table}[!hbtp]
\begin{center}
\begin{tabular}{| c | c | c | c |}
    \hline
    & $\kappa$ [kPa] & $\mu_1$ [kPa] & $\mu_2$ [kPa] \\
    \hline
    Plasma membrane  & $39.7333$  & $0.41$ & $0.422$ \\
    Cytoplasm 			  & $39.7333$  & $0.41$ & $0.422$ \\
    Nuclear envelope & $239.989$  & $2.41$ & $2.422$ \\
    Nucleoplasm 	 	  & $239.989$  & $2.41$ & $2.422$ \\
    Nucleolus 			  & $719.967$  & $7.23$ & $7.266$ \\
    ECM			 			  & $248.333$	& $5.0$	 & $5.0$   \\
    \hline
\end{tabular}
\caption{Set of constitutive parameters (bulk modulus $\kappa$ and shear moduli $\mu_1$ and $\mu_2$) used in the eigenfrequency analyses.}
\label{Tab:Parameters}
\end{center}
\end{table}

\subsection{Verification against analytical solutions}
\label{Sec:Analytical}

Based on the elastic model described in the foregoing, the remainder of the paper is concerned with the computation of the normal modes of vibration of healthy and cancerous cells using finite elements. To this end, we begin by assessing the accuracy of the finite element model used in subsequent calculations by means of comparisons to exact solutions. We consider a single spherical free-standing cell and compare numerically computed eigenmodes with the analytical solution of \cite{Kochmann:2012} for a free-standing elastic sphere with an elastic spherical inclusion.

In the harmonic range, the finite-element discretization of the model leads to the standard symmetric linear eigenvalue problem
\begin{equation}\label{Eq:Det}
    \left(\mbs{K}-\omega^2\mbs{M}\right)
    \hat{\mbs{U}}
    =
    \mbs{0} ,
\end{equation}
where $\mbs{K}$ and $\mbs{M}$ are the stiffness and mass matrices, respectively, $\omega$ is an eigenfrequency of the system and $\hat{\mbs{U}}$ is the corresponding eigenvector, subject to the normalization condition
\begin{equation}\label{E45dLS}
    \hat{\mbs{U}}^T \mbs{M} \hat{\mbs{U}} = 1 .
\end{equation}
For the spherical geometry under consideration, the modal analysis has been carried out  analytically in closed form by \cite{Kochmann:2012}. For the homogeneous sphere, they find that the natural frequencies $\omega_i$ follow as the roots of function
\begin{equation}
	f(y) = \tan{y} - \frac{y}{1-ky^2},
\end{equation}
with
\begin{equation}\label{Eq:Analytica_Solid}
	y
    =
    \sqrt{\frac{\rho\omega^2b^2}{\lambda+2\mu}}
    \quad \text{and} \quad
    k
    =
    \frac{\lambda+2\mu}{4\mu} ,
\end{equation}
where $b$ is the outer radius and $\lambda$ and $\mu$ and the Lam\'e constants. Furthermore, \cite{Kochmann:2012} report analytical solutions for an isotropic linear-elastic spherical inclusion of radius $a$, moduli $\lambda_1$ and $\mu_1$, within a concentric isotropic linear-elastic coating of uniform thickness of outer radius $b$, moduli $\lambda_2$ and $\mu_2$. In this case, the eigenfrequencies $\omega_i$ follow from the characteristic equation $\det{\mbs{A}}=0$, where
\begin{equation}
\begin{split}
    &
	A_{11}
    =
    \frac{\cos(j x)}{j x}
    -
    \frac{\sin(j x)}{j^2x^2} ,
    \\ &
	A_{12}
    =
    \frac{\sin(j k x)}{j^2k^2x^2}
    -
    \frac{\cos(j k x)}{j k x} ,
    \\ &
    A_{12}
    =
	-
    \frac{\cos(j k x)}{j^2k^2x^2}
    -
    \frac{\sin(j k x)}{j k x} ,
    \\ &
    A_{21} = 0 ,
    \\ &
    A_{21}
    =
	4 k_2^2 \frac{\cos(j k)}{j^2 k^2}
    +
    \left(\frac{1}{j k} - \frac{4k_2^2}{j^3 k^3}\right)\sin(j k) ,
    \\ &
    A_{23}
    =
    4 k_2^2 \frac{\sin(j k)}{j^2 k^2}
    -
    \left(\frac{1}{j k} - \frac{4k_2^2}{j^3 k^3}\right)
    \cos(j k) ,
    \\ &
    A_{31}
    =
	-
      k
    \left(
        4 k_1^2 \frac{\cos(j x)}{j^2 x^2}
        +
        \left(\frac{1}{j x}-\frac{4k_1^2}{j^3 x^3}\right)
        \sin(j x)
    \right) ,
    \\ &
    A_{32}
    =
	4 k_2^2 \frac{\cos(j k x)}{j^2 k^2 x^2}
    +
    \left(\frac{1}{j k x}
    -
    \frac{4k_2^2}{j^3 k^3 x^3}\right)\sin(j k x) ,
    \\ &
    A_{33}
    =
	4k_2^2
    \frac{\sin(j k x)}{j^2 k^2 x^2}
    -
    \left(\frac{1}{j k x} - \frac{4k_2^2}{j^3 k^3 x^3}\right)
    \cos(j k x) ,
\end{split}
\end{equation}
with dimensionless quantities
\begin{align}
	k_1 &= \sqrt{\frac{\mu^I}{\lambda^I+2\mu^I}}, \quad
	k_2 = \sqrt{\frac{\mu^{II}}{\lambda^{II}+2\mu^{II}}}, \quad \nonumber \\
	k &= \sqrt{\frac{\lambda^I+2\mu^I}{\lambda^{II}+2\mu^{II}}}, \quad
	j = \sqrt{\frac{\rho\omega^2 b^2}{\lambda^I+2\mu^I}}
\label{Eq:Analytical_Inclusion}
\end{align}
and with $x = a/b$.

\begin{table}[!hbtp]
\begin{center}
\begin{tabular}{| c | c | c | c | c |}
  \hline
 \textit{Constitutive parameters} & $\kappa_1$ [Pa] & $\mu_1$ [Pa] & $\kappa_2$ [Pa] & $\mu_2$ [Pa] \\
\hline
Solid sphere & $1.0$  & $1.0$ & $1.0$ & $1.0$ \\
Spherical inclusion & $1.0$  & $0.1$ & $1.0$ & $0.1$ \\
\hline
\textit{Material/ geometric parameters} & $\rho_1$ [kg/m$^3$]& $\rho_2$ [kg/m$^3$] & $r_{inner}$ [m] & $r_{outer}$ [m] \\
\hline
Solid sphere & $10^{-3}$ & $10^{-3}$ & $3$ & $30$ \\
Spherical inclusion & $10^{-3}$ & $10^{-3}$ & $3$ & $30$ \\
\hline
\end{tabular}
\caption{Set of geometric-, material- and constitutive parameters used in analytical eigenfrequency calculations and finite-element analysis.}
\label{Tab:Analytical_Parameters}
\end{center}
\end{table}

\begin{table}[!hbtp]
\begin{center}
\begin{tabular}{| c | c | c |}
\hline
 & $\omega_{lowest}$ [rad/s] (I) & $\omega_{lowest}$ [rad/s] (II) \\
\hline
Analytical solution & 3.72394 & 1.17402 \\
Finite element analysis & 3.71717 & 1.17547 \\
Relative error (\%) & 0.181797 & 0.123507 \\
\hline
\end{tabular}
\caption{Lowest radial eigenfrequency from analytical solution and finite-element analysis for the cases of a solid sphere (I) and a spherical matrix with inclusion (II).}
\label{Tab:Analytical_Solution}
\end{center}
\end{table}

Table~\ref{Tab:Analytical_Solution} shows a comparison between analytical and finite-element values of the fundamental frequency of a solid sphere and a sphere with a high-contrast spherical inclusion for the particular choice of parameters listed in Table~\ref{Tab:Analytical_Parameters}. The finite-element values correspond to a mesh of $\approx$ $15,000$ linear tetrahedral elements, representative of the meshes used in subsequent calculations. As may be seen from the table, the finite-element calculations may be expected to result in frequencies accurate to $\sim 10^{-3}$ relative error.

\subsection{Eigenfrequencies and eigenmodes} \label{Sec:Frequencies}

\begin{figure}[]
\begin{center}
    \includegraphics[width=0.8\textwidth]{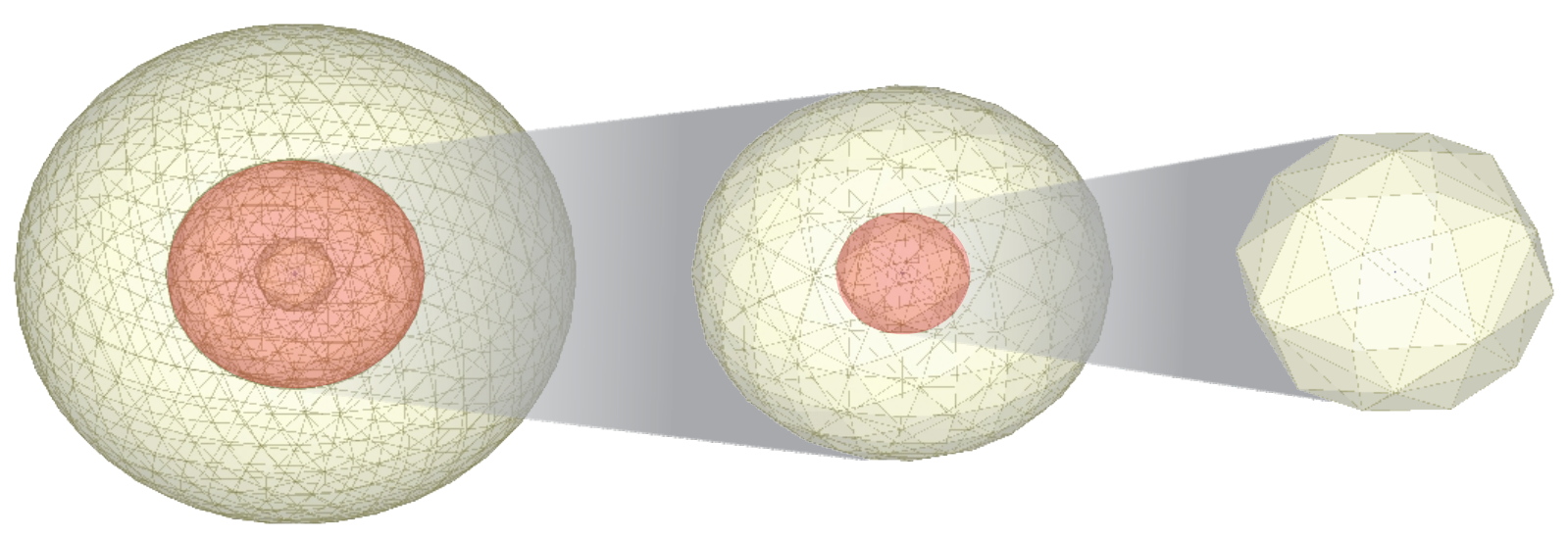}
\caption{Finite-element mesh of the plasma membrane, cytoplasm, nuclear envelope, nucleoplasm, and nucleolus (for the purpose of visualization, the extracellular matrix is omitted).}
\label{Fig:Mesh}
\end{center}
\end{figure}

{In order to obtain a first estimate of the spectral gap between cancerous and healthy cells, we consider the eigenvalue problem of single free-standing cells. To this end, cytoplasm, nucleoplasm and nucleolus are discretized using linear tetrahedral elements, while linear triangular membrane elements are used for the plasma membrane and nuclear envelope. A typical finite element mesh for a cell geometry with a ratio of $n/c=1$ and a total of $40,349$ elements is shown in Figure~\ref{Fig:Mesh}. For the calculation of eigenfrequencies, meshes containing $\sim 16,000$ elements are used for geometries ranging from a ratio of $n/c=1$ to $n/c=2$. In addition to the accuracy assessment of Section~\ref{Sec:Analytical}, we have assessed the convergence of the free-standing cell finite-element model by considering five different meshes of $2,171$, $3,596$, $8,608$, $11,121$, and $15,215$ elements. From this analysis, we find that the accuracy in the lowest eigenfrequency for the finest mesh is of the order of $0.2\%$, which we deem sufficient for present purposes.} All cell constituents are modeled by means of the hyperelastic Mooney-Rivlin model described in Section~\ref{Sec:Geometry_Material}, with the materials constants of Table~\ref{Tab:Parameters}.

\begin{figure}[]
\begin{center}
    \includegraphics[width=0.7\textwidth]{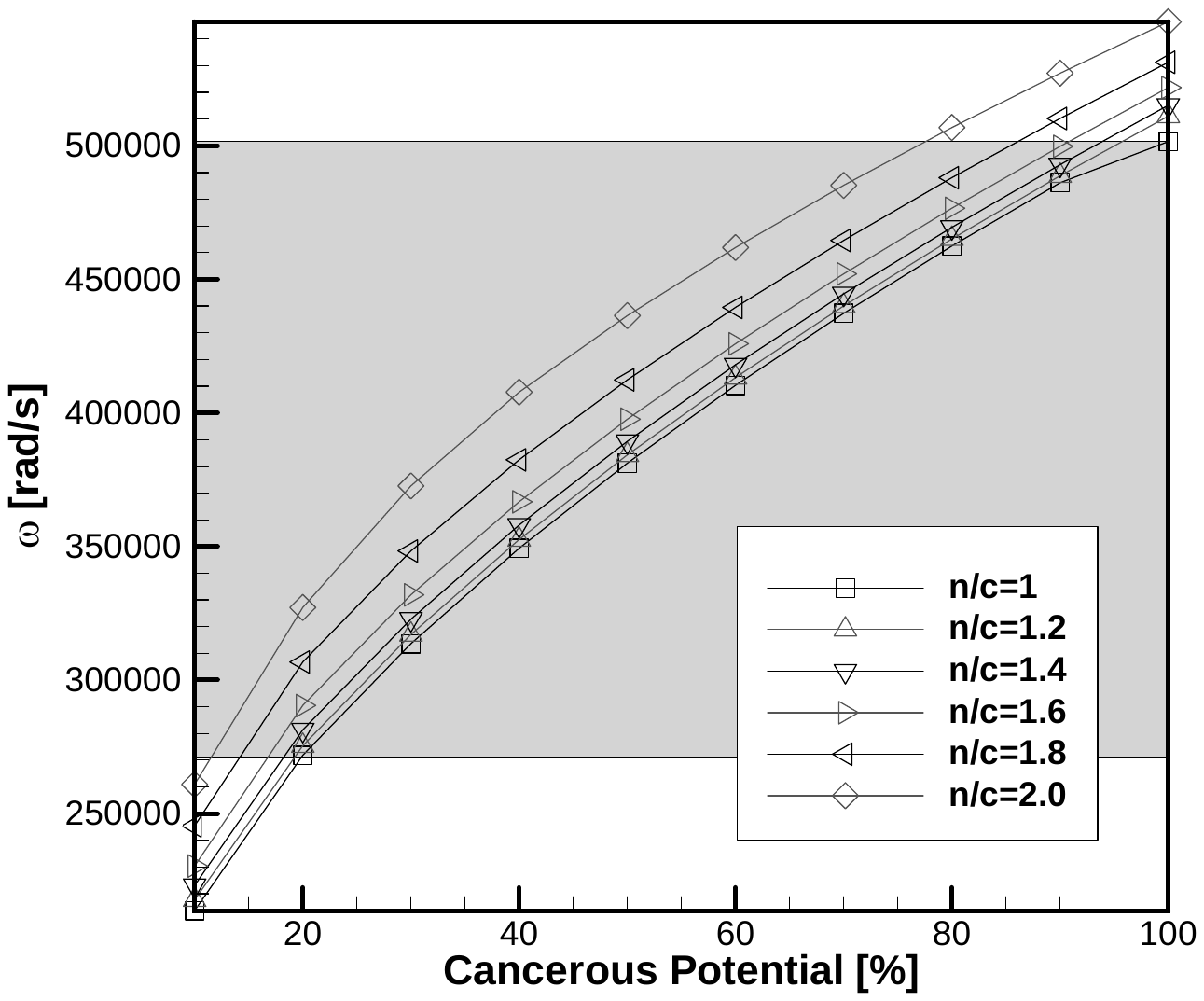}
\caption{Lowest eigenfrequency for varying cell stiffness and increasing nucleolus/nucleoplasm-to-cytoplasm volume ratios $n/c$.}
\label{Fig:Frequencies}
\end{center}
\end{figure}

Figure~\ref{Fig:Frequencies} shows the calculated lowest eigenfrequency, rigid-body modes excluded, for different cell geometries and varying material properties. In the calculations, the nucleolus/nucleoplasm-to-cytoplasm volume ratio is increased incrementally in the range of $n/c=1.0$ to $n/c=2.0$, resulting in six different test geometries. Furthermore, material properties are varied within a pathophysiologically relevant range, whereby a value of $100\%$ cancerous potential corresponds to values presented in Table~\ref{Tab:Parameters}.

Since cancerous cells are more than $80\%$ softer than healthy cells \citep{Cross:2007}, we vary the elastic moduli in Table~\ref{Tab:Parameters}, with full values representing cancerous cells and increased moduli representing healthy cells. In addition, decreased elastic moduli of the extracellular matrix (ECM) are expected in healthy tissues \citep{Schrader:2010}. Figure~\ref{Fig:Frequencies} summarizes the calculated lowest eigenfrequency for different percentages of the parameter values presented in Table~\ref{Tab:Parameters}. Thus, a cancerous potential of $80\%$ reflects an increase in elastic moduli of $20\%$ for material parameters of the different cell constituents with a simultaneous decrease in elastic moduli of the ECM by $20\%$. The shaded area in Figure~\ref{Fig:Frequencies} illustrates a typical gap in the lowest natrual frequency for a nucleolus/nucleoplasm-to-cytoplasm volume ratio of $n/c=1.0$, with $\omega=501,576$~rad/s for cancerous cells and $\omega=271,764$~rad/s for a reduction in cancerous potential by $80\%$, the expected value for healthy cells \citep{Cross:2007}). An even higher spectral gap is recorded by additionally taking the growth in nucleolus/nucleoplasm-to-cytoplasm volume ratio into account, as experimentally observed in cancerous cells \citep{Berman:2011}.

A more detailed comparison of the spectra of free-standing healthy and cancerous cells, corresponding to cancerous potentials of $20\%$ and $100\%$, respectively, is presented in Table~\ref{Tab:Frequencies}, which collects the computed lowest ten eigenfrequencies for a cell geometry with volume ratio $n/c=1.0$. From this table we observe that free-standing cancerous cells have a ground eigenfrequency of the order of $500,000$ rad/s, whereas free-standing healthy cells have a ground eigenfrequency of the order of $270,000$ rad/s, or a healthy-to-cancerous spectral gap of the order of $230,000$ rad/s. In addition, their higher eigenfrequencies overlap with the ground eigenfrequency of cancerous cells. Therefore, special attention is required to examine whether or not excitation of cancerous cells might trigger healthy cells to resonate. Indeed, figures of merit other than natural frequency, including growth rates of resonant modes and energy absorption, may also be expected to play an important role in differentiating the response of cancerous and healthy cells. These additional figures of merit are investigated in Section~\ref{Sec:Transient}.

\begin{figure}[]
\begin{center}
    \includegraphics[width=0.9\textwidth]{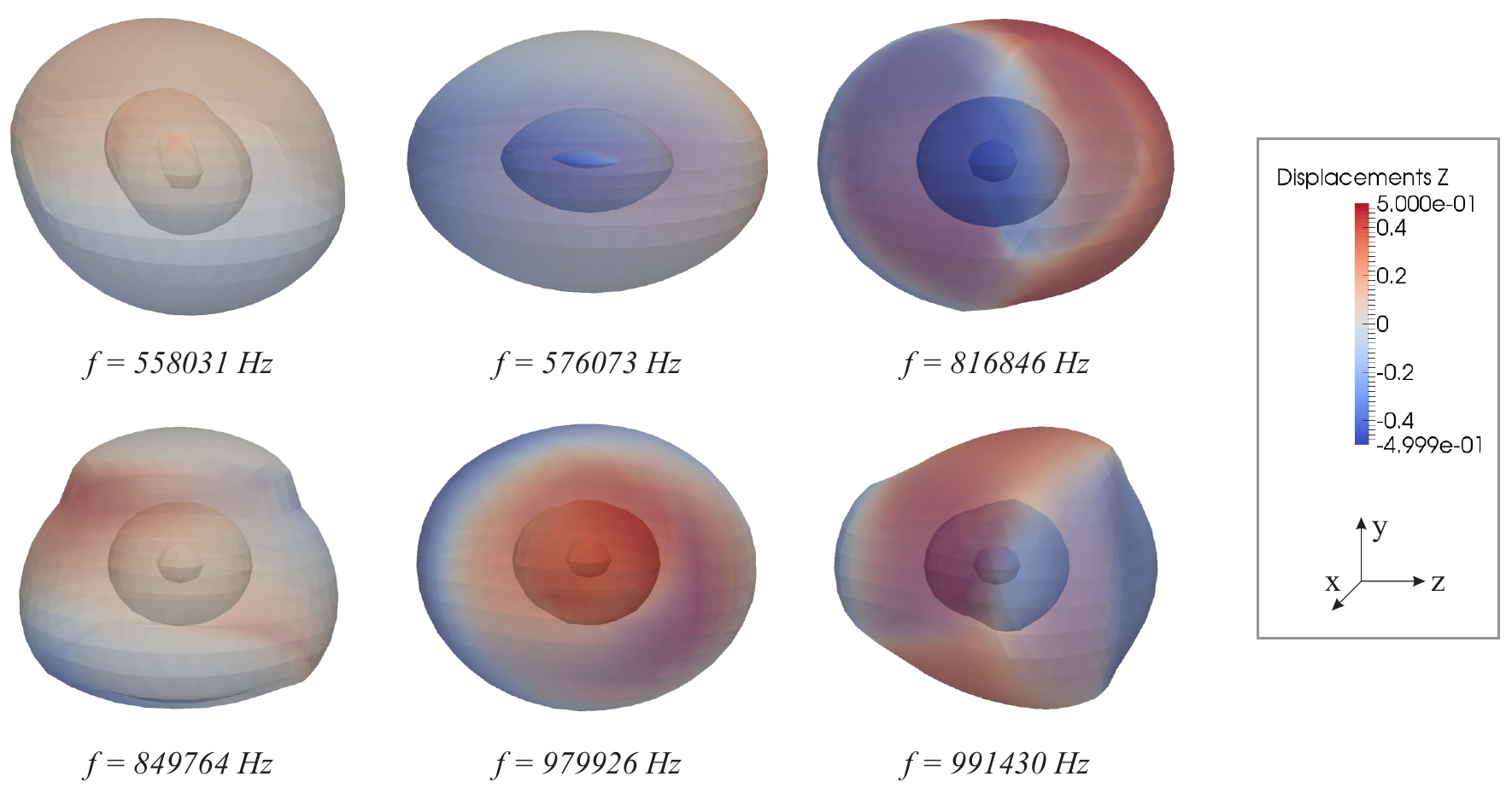}
\caption{Eigenmodes corresponding to different resonance frequencies for a ratio of $n/c=1.0$ and a cancerous potential of $100\%$.}
\label{Fig:Modes}
\end{center}
\end{figure}

The eigenmodes corresponding to different resonance frequencies for a ratio of $n/c=1.0$ and a cancerous potential of $100\%$ are shown in Figure~\ref{Fig:Modes}. It may be noted from the figure how each mode represents different characteristic deformation mechanisms of the various cell constituents. Knowledge of the precise modal shape may therefore help to target lysis of specific cell components. Thus, shear deformation may be expected to dominate at a frequency of $558,031$ rad/s, whereas volumetric deformations may be expected to be dominant at $576,073$ rad/s. These differences in deformation mode open the way for targeting specific cell constituents for lysis, such as the plasma membrane at a frequency of $816,846$ rad/s.

\begin{table}[!hbtp]
\begin{center}
\begin{tabular}{| c | c | c | c | c | c | c |}
\hline
 & $\omega_1$ [rad/s] & $\omega_2$ [rad/s] & $\omega_3$ [rad/s]  & $\omega_4$ [rad/s]  & $\omega_5$ [rad/s] \\
\hline
Cancerous & 501576 & 502250 & 508795 & 532132 & 537569 \\
Healthy & 271764 & 274141 & 364259 & 364482 & 367413 \\
\hline
 & $\omega_6$ [rad/s] & $\omega_7$ [rad/s] & $\omega_8$ [rad/s]  & $\omega_9$ [rad/s]  & $\omega_{10}$ [rad/s] \\
\hline
Cancerous & 538512 & 557291 & 667107 & 678287 & 678771 \\
Healthy & 375570 & 376000 & 380063 & 424226 & 425327 \\
\hline
\end{tabular}
\caption{Comparison of the lowest ten eigenfrequencies for a cell geometry with a ratio of $n/c=1.0$ and a cancerous potential of $100\%$ (cancerous) versus a cancerous potential of $20\%$ (healthy) obtained from a free vibration analysis.}
\label{Tab:Frequencies}
\end{center}
\end{table}

\subsection{Bloch wave analysis} \label{Sec:Bloch}

\begin{figure}[]
\begin{center}
    \includegraphics[width=0.72\textwidth]{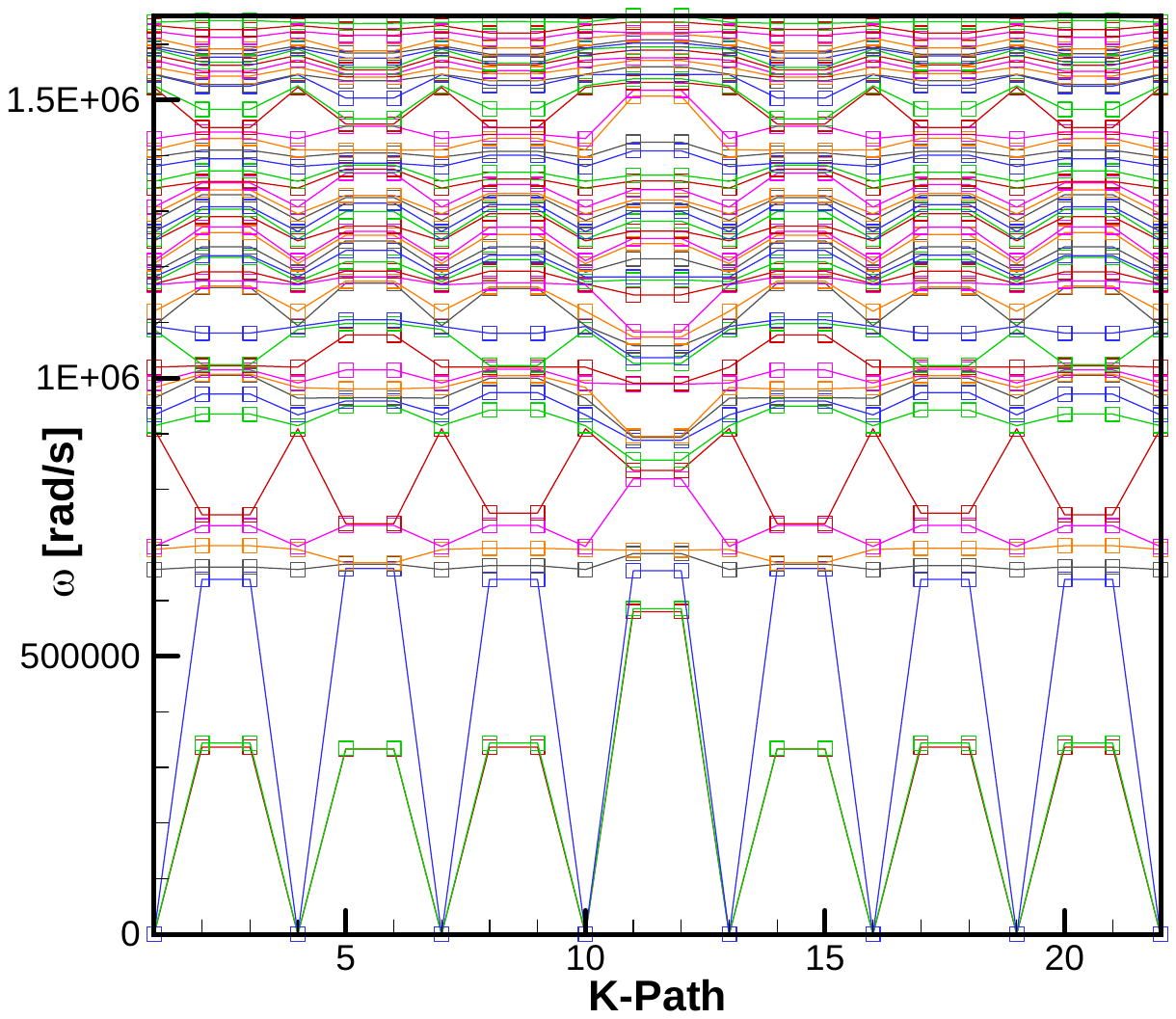}
\caption{{Disperison relations for a cancerous cell, whereby the k-path is traversed with a refinement of two points between neighboring nodes.}}
\label{Fig:Disp_Cancerous}
\end{center}
\end{figure}

\begin{figure}[]
\begin{center}
    \includegraphics[width=0.72\textwidth]{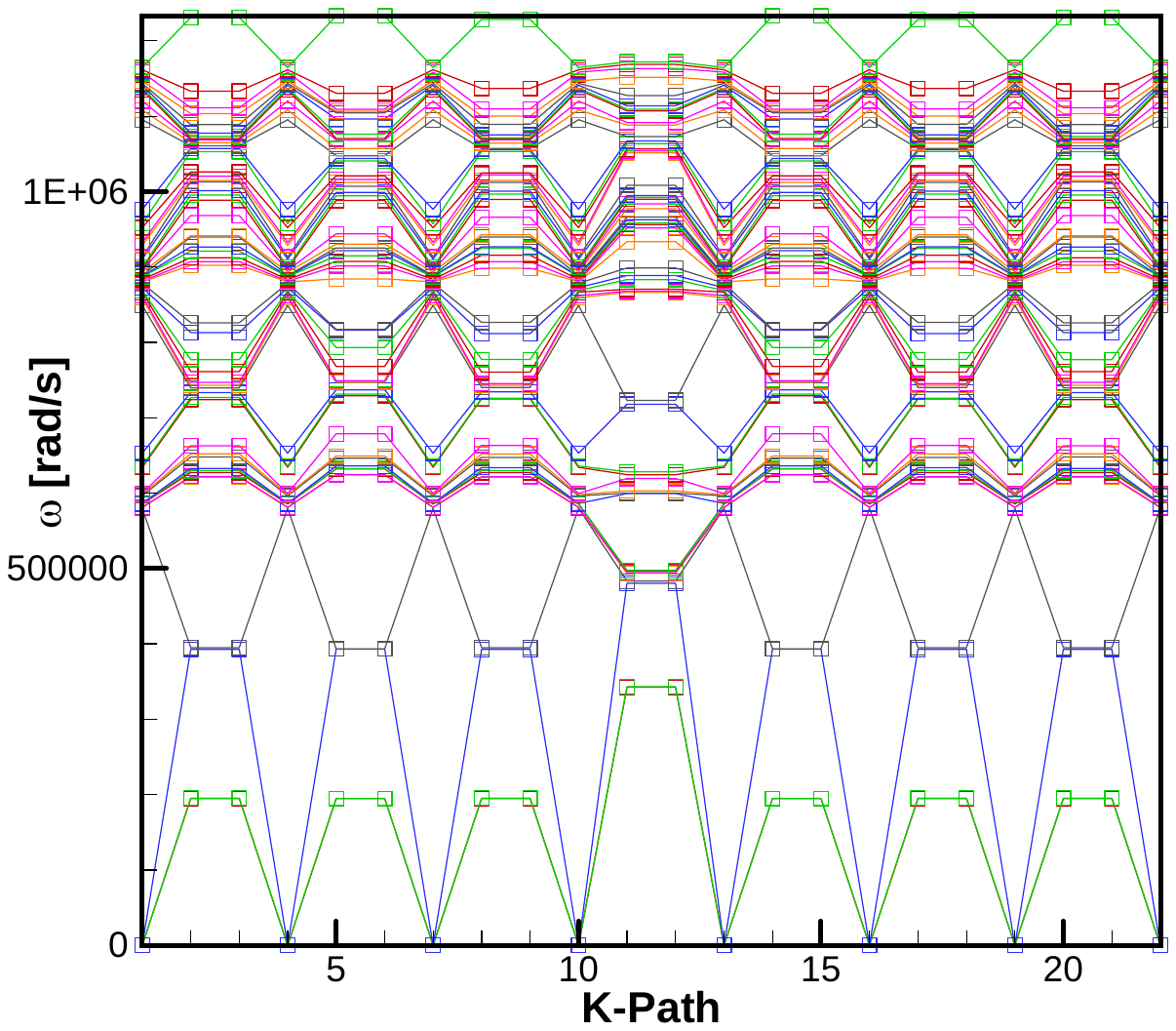}
\caption{{Disperison relations for a healthy cell, whereby the k-path is traversed with a refinement of two points between neighboring nodes.}}
\label{Fig:Disp_Healthy}
\end{center}
\end{figure}

{The preceding spectral analysis for a free-standing cell can be extended to a tissue consisting of a periodic arrangement of cells embedded into an extracellular matrix. In this case, the analysis can be carried out by recourse to standard Bloch wave theory. Within this framework, the displacement field is assumed to be of the form
\begin{equation}\label{74RQg5}
	\mbs{u}(\mbs{x})
    =
    \hat{\mbs{u}}(\mbs{x})
    e^{i\mbs{k}\cdot\mbs{x}},
\end{equation}
where $\mbs{k}$ is the wave vector of the applied harmonic excitation and the new unknown displacement field $\hat{\mbs{u}}(\mbs{x})$ is defined within the periodic cell \citep{Bloch:1929}. By periodicity, the values of wave vector $\mbs{k}$ can be restricted to the Brillouin zone of the periodic lattice. Substitution of representation (\ref{74RQg5}) into the equations of motion results in a $\mbs{k}$-dependent eigenvalue problem. The corresponding eigenfrequencies $\omega_i(\mbs{k})$ define the dispersion relations of the tissue.  Details of implementation of Bloch-wave theory in elasticity and finite-element analysis may be found in \cite{Kroedel:2013}.

\begin{figure}[!h]
\begin{center}
    \includegraphics[width=0.8\textwidth]{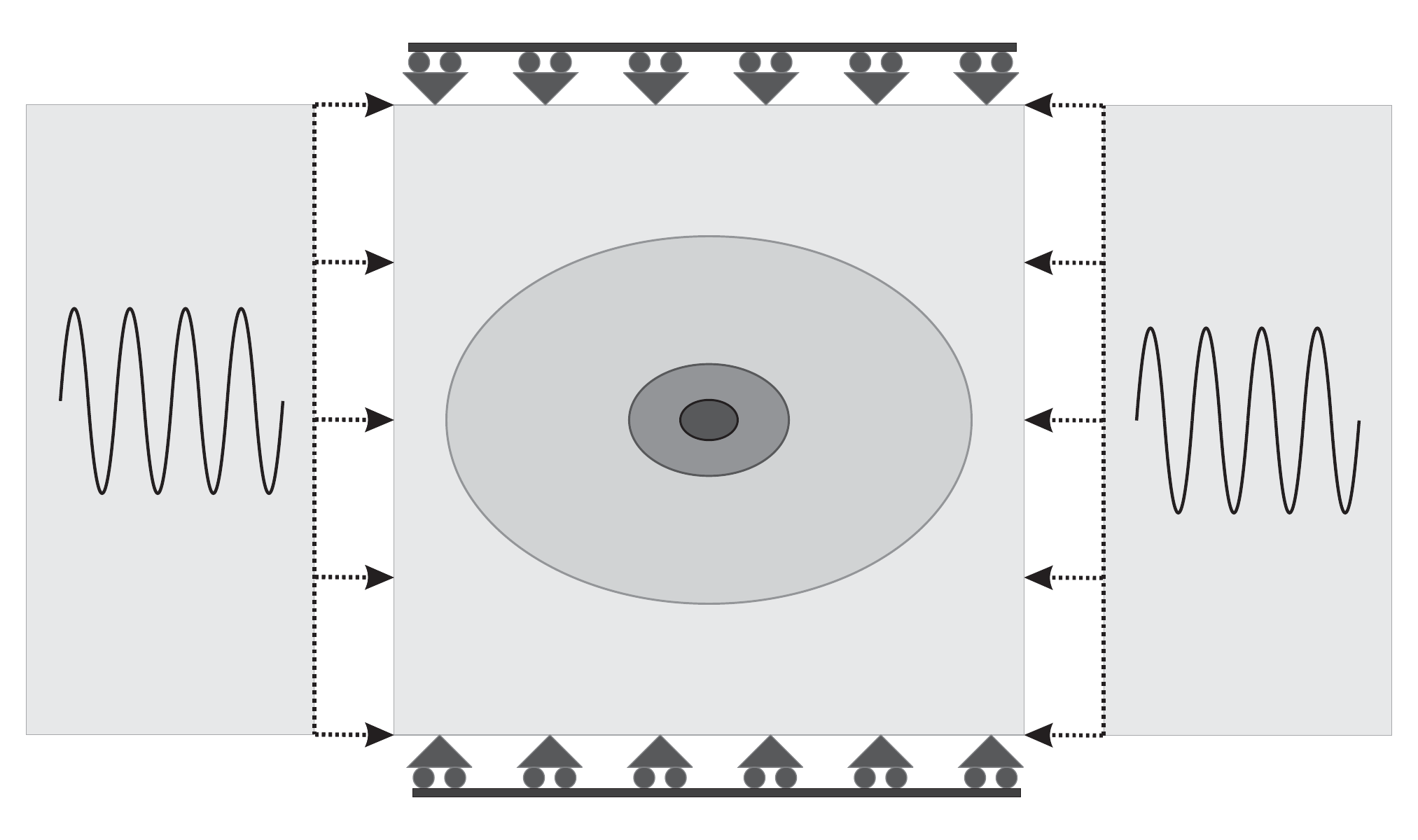}
\caption{Displacement elastodynamic boundary value problem with applied harmonic excitation.}
\label{Fig:BVP}
\end{center}
\end{figure}

\begin{figure}[]
\begin{center}
    \includegraphics[width=0.8\textwidth]{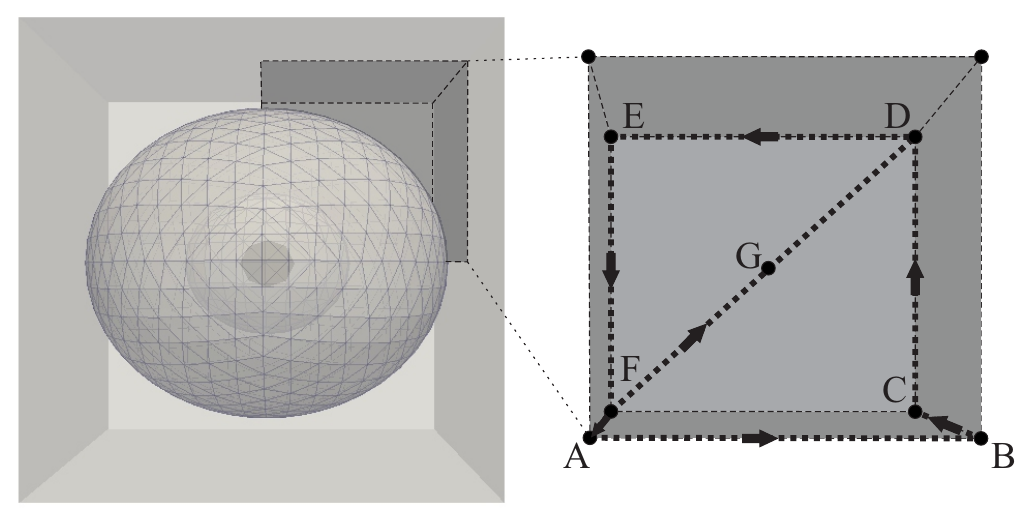}
\caption{First irreducible Brillouin zone and chosen $k$-path.}
\label{Fig:Zone}
\end{center}
\end{figure}

In the present analysis, we consider a cubic unit cell of size $a=15$\,$\mu$m and the finite element discretization shown in Figure~\ref{Fig:Mesh}. The extracellular matrix (ECM), not shown in the figure, is also discretized into finite elements. Figure~\ref{Fig:Zone} shows the first irreducible Brillouin zone, which is itself a cube of size $2\pi/a$. In order to visualize the dispersion relations, we choose the $k$-path along the edges and specific symmetry lines of the Brillouin zone also shown in Figure~\ref{Fig:Zone}. The path allows for the elliptical shape of the cells, with only one symmetry axis. The computed dispersion relations for both cancerous and healthy cells then follow as shown in Figures~\ref{Fig:Disp_Cancerous} and~\ref{Fig:Disp_Healthy}, respectively, for the lowest $50$ eigenfrequencies. Similarly to the calculations presented in Section~\ref{Sec:Frequencies} for free-standing cells, the lowest eigenfrequencies of the healthy tissue are shifted uniformly towards lower values with respect to the eigenfrequencies of the cancerous tissue, with significant spectral gaps of the order of $200,000$ rad/s between the two.}


We recall that the computed ground eigenfrequency of free-standing cancerous cells is of the order of $\omega \sim 500,000$ rad/s. In addition, from the properties of Table~\ref{Tab:Parameters} we may expect a cancerous-tissue shear sound speed of the order of $c=\sqrt\frac{{\mu}}{{\rho}}=\sqrt{\frac{3\kappa(1-2\nu)}{2\rho(1+\nu)}}\sim 0.8$\,m/s (cytoplasm) to $c\sim 7.2$\,m/s (nucleolus). Therefore, at resonance, the corresponding wave number of the applied harmonic excitation is of the order of $k \sim \omega/c \sim 725,000$\,rad/m (cytoplasm) and $k\sim 69,444$\,rad/m (nucleolus) or, correspondingly, a wavelength of the order of $\lambda \sim 2\pi/k \sim 10^{-5}$ m (cytoplasm) and $\lambda\sim 9\cdot 10^{-5}$ m (nucleolus), which is larger than a typical cell size. It thus follows that the regime of interest here is the long-wavelength regime, corresponding to the limit of $k\to 0$ in the preceeding Bloch-wave analysis. Consequently, in the remainder of the paper we restrict attention to that limit. The corresponding boundary value problem takes the form sketched in Figure~\ref{Fig:BVP} and consists of the standard displacement elastodynamic boundary value problem with harmonic displacement boundary conditions applied directly to the boundary.

\begin{table}[!hbtp]
\begin{center}
\begin{tabular}{| c || c | c | c | c | c |}
\hline
$\omega_n$ [rad/s] & 501576 & 502250 & 532132 & 537569 \\
$r_{n,cancerous}\,||\hat{U}_n||$ [$\frac{\mu m}{s}$] & $8.862\cdot 10^6$ & $9.179\cdot 10^6$ & $-3.898\cdot 10^8$ & $-2.863\cdot 10^7$ \\
\hline
$\omega_n$ [rad/s] & 496165 & 496165 & 519049 & 545277 \\
$r_{n,healthy}\,||\hat{U}_n||$ [$\frac{\mu m}{s}$] & $-3.882\cdot 10^6$ & $-3.882\cdot 10^6$ & $-2.032\cdot 10^6$ & $-0.335\cdot 10^6$ \\
\hline
\end{tabular}
\caption{Comparison of growth rate ratios $r_{n,cancerous}$ and $r_{n,healthy}$ with $r_n=\frac{F_n}{2\omega_n}$. Shown are values corresponding to the lowest four cancerous eigenfrequencies $\omega_{n,cancerous}$ and their related closest healthy eigenfrequencies $\omega_{n,healthy}$.}
\label{Tab:Ratios}
\end{center}
\end{table}

\begin{figure}[h]
\begin{center}
    \subfigure[]{
    \includegraphics[width=0.48\textwidth]{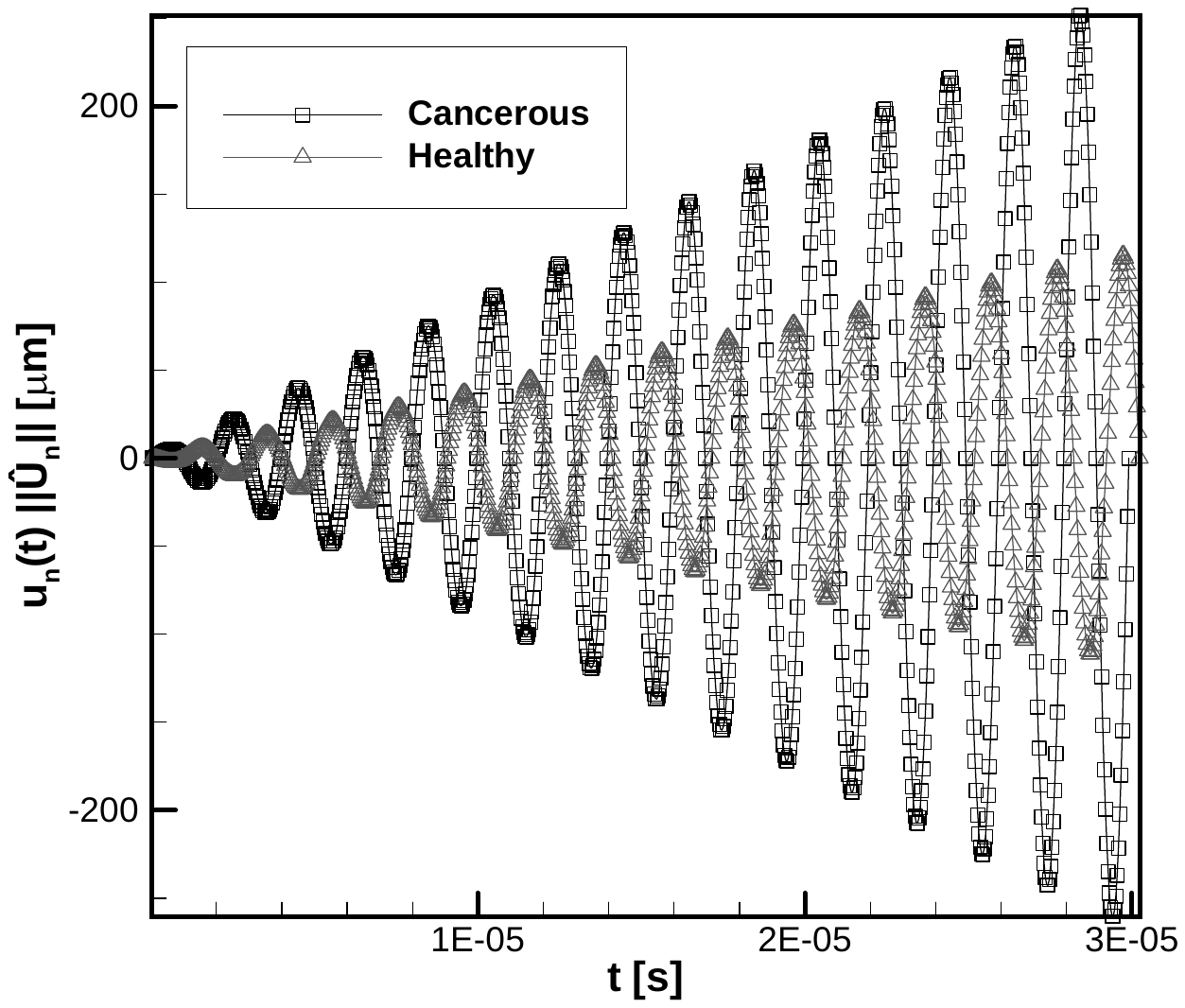}}
    \subfigure[]{
    \includegraphics[width=0.48\textwidth]{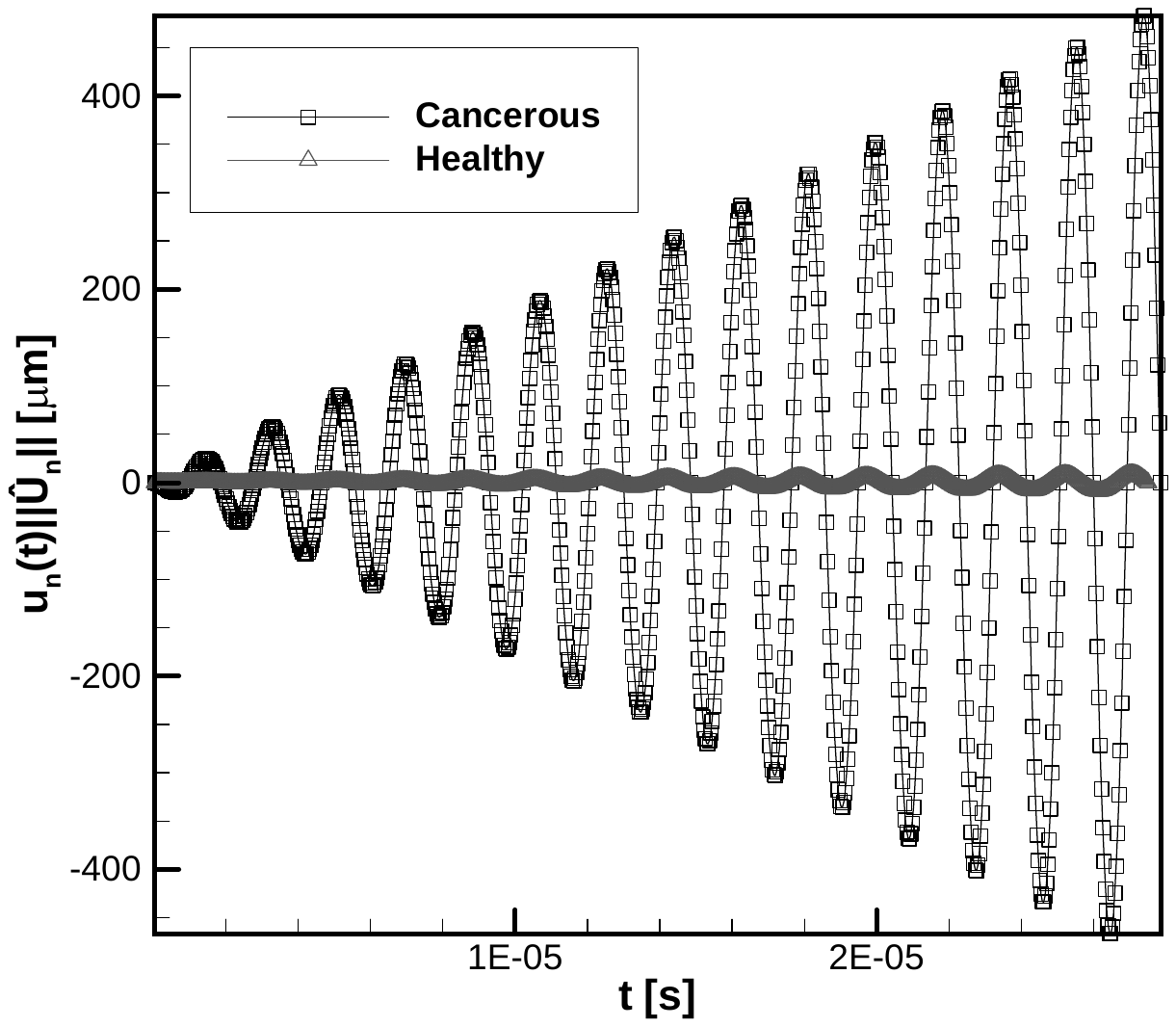}}
\caption{{Comparison of $u_n$ during transient response simulations in the linearized kinematics framework for a cancerous cell (with a cancerous potential of $100\%$) excited at one of its resonance frequencies $\omega_{c}$ and a healthy cell (with a cancerous potential of $20\%$) excited at an eigenfrequency $\omega_{h}$ which is closest to $\omega_{c}$.}}
\label{Fig:Fn}
\end{center}
\end{figure}

\subsection{Resonant growth rates}

The spectral gap, or gap in the lowest eigenfrequencies, between healthy and cancerous cells and tissues provides a first hint of sharp differences in the response of healthy and cancerous tissue to harmonic excitation. In particular, the preceding analysis shows that the fundamental frequencies of the cancerous tissue may be in close proximity to eigenfrequencies of the healty tissue, which appears to undermine the objective of selective excitation of the cancerous tissue. However, a complete picture requires consideration of the relative energy absorption characteristics and growth rates of resonant modes. To this end, we consider the modal decomposition of the displacement field
\begin{equation}
	\mbs{U}(t) = \sum_{n=1}^N u_n(t) \hat{\mbs{U}}_n ,
\end{equation}
where $(\hat{\mbs{U}}_n)_{n=1}^N$ are eigenvectors obeying the orthogonality and normalization condition (\ref{E45dLS}) and $(u_n(t))_{n=1}^N$ are time-dependent modal amplitudes obeying the modal equations of motion
\begin{equation}\label{Kc3V7h}
	\ddot{u}_n(t) + \omega^2_n u_n(t)
    =
    \hat{\mbs{U}}^T_n \mbs{F}_{ext}(t)
    =
    F_n(t) .
\end{equation}
In this equation, $\omega_n$ is the corresponding eigenfrequency, $\mbs{F}_{ext}(t)$ is the external force vector and $F_n(t)$ is the corresponding modal force. For a harmonic excitation of frequency $\omega_{ext}$, eq.~(\ref{Kc3V7h}) further specializes to
\begin{equation}
	\ddot{u}_n(t) + \omega_n ^2 u_n(t) = F_n \cos{\omega_{ext} t},
\end{equation}
where now $F_n$ is a constant modal force amplitude. At resonance, $\omega_{ext} = \omega_n$, the amplitude of the transient solution starting from quiescent conditions grows linearly in time and the transient solution follows as
\begin{equation}
    u_n(t)
    =
    \frac{F_n}{2\omega_n}t\sin({\omega_nt}) .
\end{equation}
We thus conclude that the growth rate of resonant modes is
\begin{equation}
    r_n = \frac{F_n}{2\omega_n} .
\end{equation}

Figure~\ref{Fig:Fn} shows the growth properties of $r_n$ for two different cases. In the first case, a cancerous cell is excited at its resonant frequency of $\omega_c=501,576$\,rad/s, whereas the healthy cell is excited at its closest resonance frequency of $\omega_h=496,165$\,rad/s. In the second case, eigenfrequencies of $538,512$\,rad/s and $545,277$\,rad/s are investigated. The simulations reveal that the growth rate of the resonant response of the cancerous cells is much faster than that of the healthy cells, which opens a window for selectively targeting the former.

\subsection{Transient response at resonance} \label{Sec:Transient}

The preceding analysis has been carried out with a view to understanding the resonant response of cells and tissues under harmonic excitation in the harmonic range. In this section, we seek to confirm and extend the conclusions of the harmonic analysis by carrying out fully nonlinear implicit dynamics simulations of the transient response of healthy and cancerous cells under resonant harmonic excitation. In this analysis, a geometry of ratio $n/c=1$ is considered, Figure~\ref{Fig:Mesh}, together with material parameters of Table~\ref{Tab:Parameters}. We restrict attention to the long wavelength limit, i.~e. to ultrasound radiation of wavelengths larger than the cell size. In keeping with this limit, we enforce harmonic displacement boundary conditions directly as shown in Figure~\ref{Fig:Geometry} in order to mechanically excite the cell. The strength of the harmonic excitation used in the calculations is $\hat{u}_{0}=0.04$\,$\mu$m.

In the simulations, we track the transient amplification of the cell response up to failure. We assume that failure occurs when the stress in the cytoskeletal polymer network, which constitutes the structural support for cell membranes, reaches a threshold strength value. \cite{Lieleg:2009} found that the macroscopic network strength can be traced to the microscopic interaction potential of cross-linking molecules and other cytoskeletal components such as actin filaments. Here, we assume a rupture strength of the order of $30$\,Pa based on strength values of a single actin/cross-linking protein bond reported in \cite{Lieleg:2009}.

\begin{figure}[]
\begin{center}
    \includegraphics[width=1.0\textwidth]{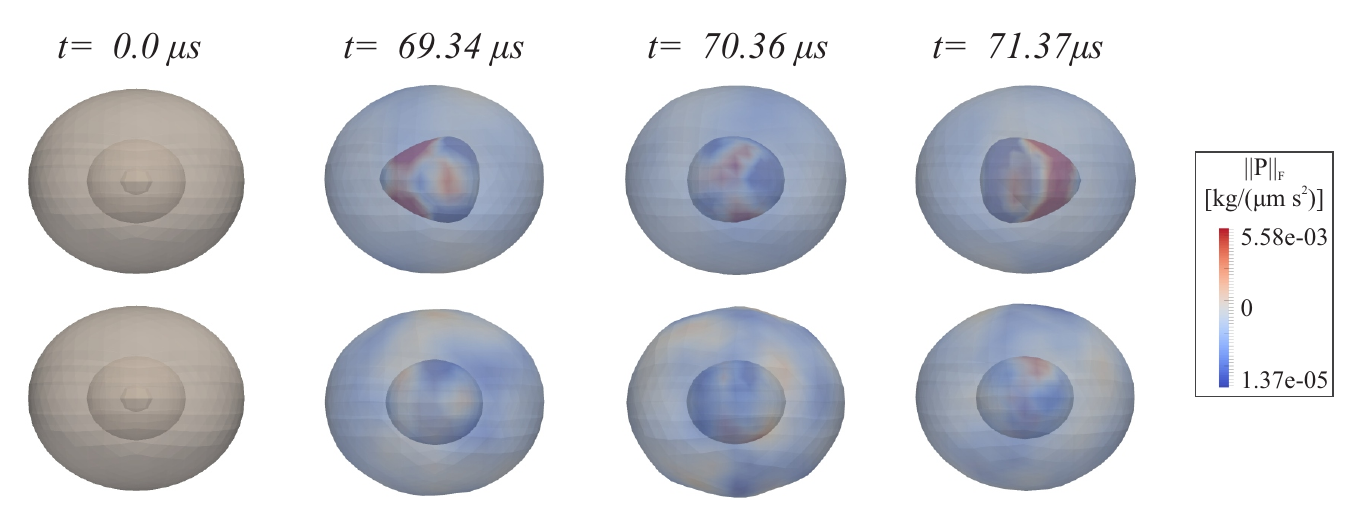}
\caption{Transient response of a cancerous cell (top) and a healthy cell (bottom) at resonance for a ratio of $n/c=1.0$ and a cancerous potential of $100\%$. Shown is the Frobenius norm of first Piola-Kirchhoff stress tensor.}
\label{Fig:Transient}
\end{center}
\end{figure}

\begin{figure}[h]
\begin{center}
    \subfigure[]{
    \includegraphics[width=0.48\textwidth]{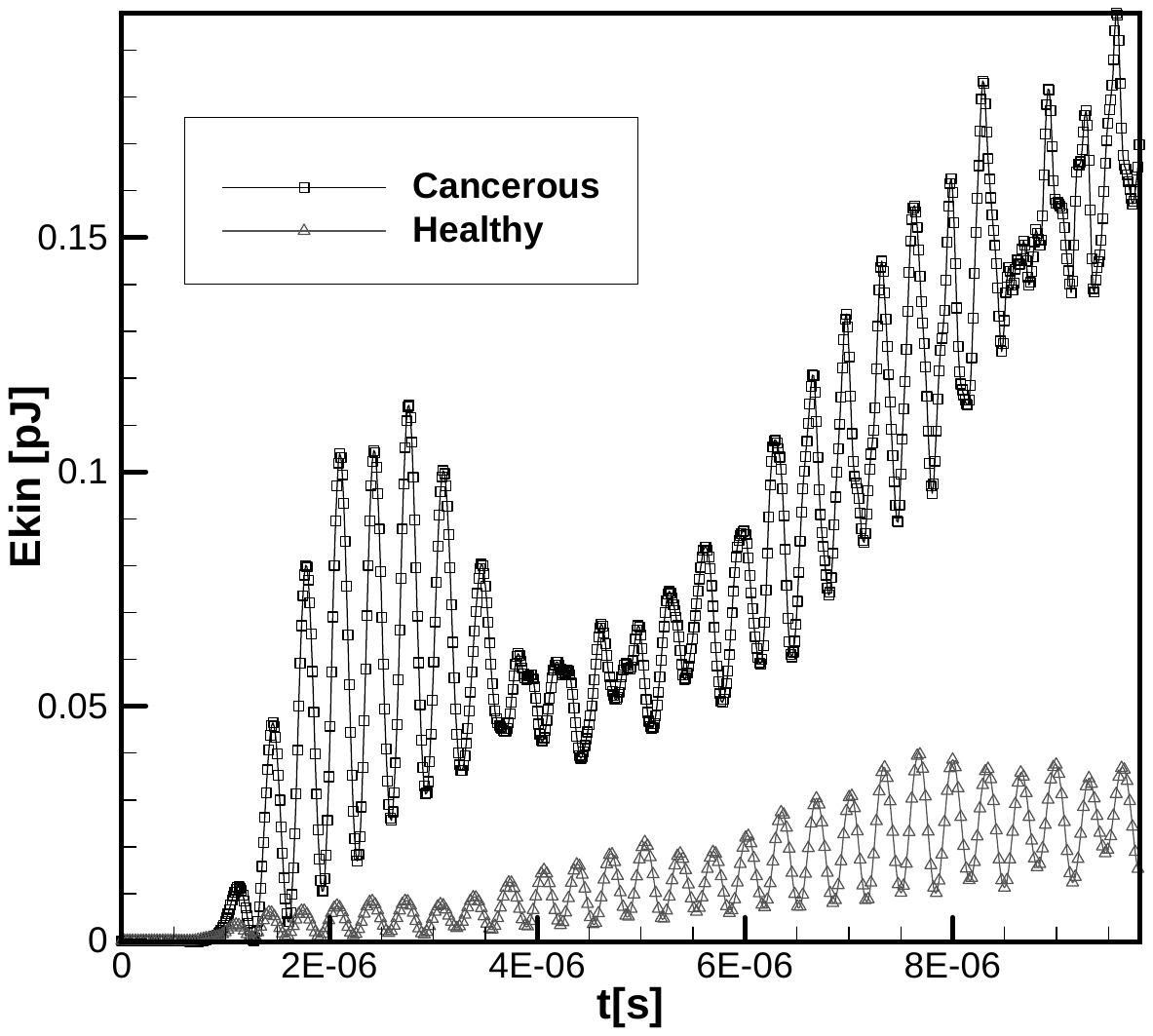}}
    \subfigure[]{
    \includegraphics[width=0.48\textwidth]{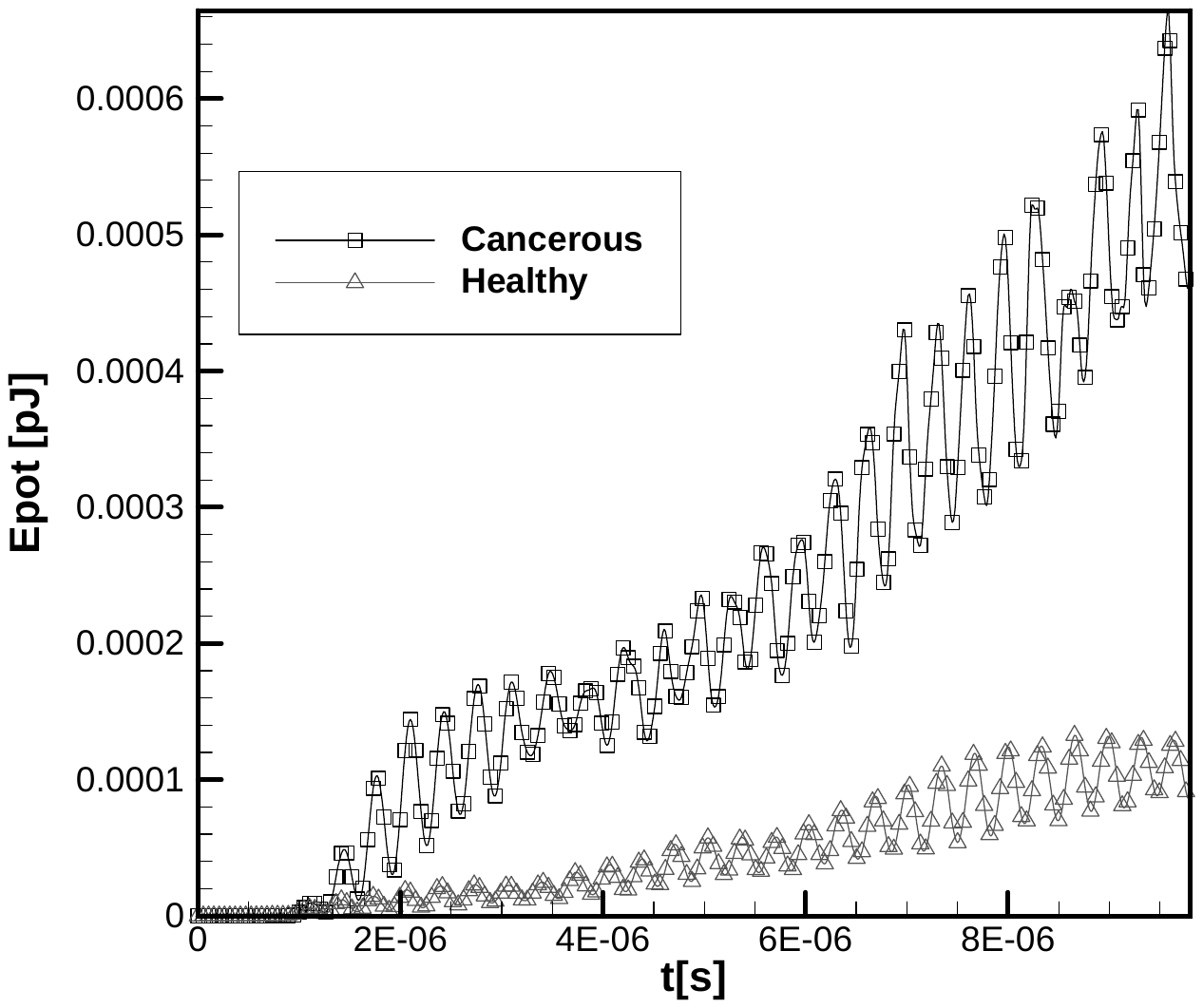}}
\caption{{Kinetic energy (left) and potential energy (right) of the nuclear envelope during excitation at resonance of both healthy and cancerous cells.}
\label{Fig:KineticPotential}}
\end{center}
\end{figure}

Figure~\ref{Fig:Transient} shows the fully-nonlinear transient response of healthy and cancerous cells at the resonant frequency of the latter. It can be seen from the figure that stresses in both the plasma membrane and nuclear envelope of the cancerous cell grow at a much faster rate than in healthy cells. For the harmonic excitation under consideration, the strength of the nuclear envelope of the cancerous cell reaches the rupture strength at time $t_{\rm lysis} \approx 71\mu$s, while, at the same time, the level of stress in the healthy cells is much lower. Figure~\ref{Fig:KineticPotential} furthermore illustrates the kinetic and potential energy of the nuclear envelope during excitation at resonance of both healthy and cancerous cells.

From transient response simulations, the energy that needs to be supplied until the point of rupture is reached is
\begin{equation}
\begin{split}
	E_{\rm lysis}
    & =
    \int_{t=0}^{t_{\rm lysis}}
    \int_{\partial\Omega}\mbs{t}\cdot\dot{\mbs{u}}\,dS\,dt
	\\ & \approx
    \sum_{i=0}^{n_{\rm lysis}}
    \sum_{j\in\partial\Omega}
    \frac{1}{2}
    \big(\mbs{F}_j(t_{i+1}) + \mbs{F}_j(t_i)\big)
    \cdot
    \big( \mbs{u}_j(t_{i+1}) - \mbs{u}_j(t_i) \big),
\end{split}
\end{equation}
where $\mbs{t}$ is the applied traction on the boundary $\partial\Omega$, $\mbs{u}$ is the displacement vector, $\mbs{F}_j(t_i)$ is the force acting on surface node $j$ at time $t_i$, and $\mbs{u}_j(t_i)$ is the corresponding displacement vector. {For a cell geometry with a ratio of $n/c=1.0$ and a cancerous potential of $100\%$, calculations give a value of $228$\,pJ for the energy per cell required for lysis.} Assuming an average cell size of 20 $\mathrm{\mu}$m, a time to lysis of 70 $\mathrm{\mu}$s and a tumor of 1 cm in size, this energy requirement translates into a power density requirement in the range of 0.8 W/cm${}^2$.

\section{Discussion and outlook}

In this study, we have presented numerical calculations that suggest that spectral gaps between hepatocellular carcinoma and healthy cells can be exploited to selectively bring the cancerous cells to lysis through the application of carefully tuned ultrasound harmonic excitation, while keeping healthy cells intact. We refer to this procedure as {\sl oncotripsy}. A normal mode analysis in the harmonic range reveals the existence of a healthy-to-cancerous spectral gap in ground frequency of the order of $230,000$ rad/s, or $36.6$ kHz. Further analysis of the growth rates of the transient response of the cells to harmonic excitation reveals that lysis of cancerous cells can be achieved without damage to healthy cells. These findings point to oncotripsy as a novel opportunity for cancer treatment via the application of carefully tuned ultrasound pulses in the frequency range of 80 kHz, duration in the range of 70 $\mathrm{\mu}$s and power density in the range of 0.8 W/cm${}^2$. This type of ultrasound actuation can be readily delivered, e.~g., by means of commercial low-frequency and low-intensity ultrasonic transducers. Evidently, the present numerical calculations serve only as preliminary evidence of the viability of oncotripsy, and further extensive laboratory studies would be required in order to confirm and refine the present findings and definitively establish the viability of the procedure.

\pagebreak
\newpage

\end{document}